%% file: sample-base.tex
\documentclass[sigplan,twocolumn]{acmart}
\renewcommand\footnotetextcopyrightpermission[1]{}
\usepackage{prp-macros}
\usepackage{amsmath,amsfonts}
\usepackage{graphicx}
\usepackage{textcomp}
\usepackage{xcolor}
\usepackage{hyperref}
\usepackage{layouts}
\usepackage{pgfplots}
\usepackage{float}
\usepackage{caption}
\usepackage{subcaption}
\usepackage{multirow}
\usepackage{booktabs}
\usepackage{indentfirst}
\usepackage{mathtools}
\usepackage{witharrows}
\usepackage[export]{adjustbox}
\usepackage{url}
\usepackage{lipsum}
\usepackage[utf8]{inputenc}
\usepackage{fancyhdr}
\usepackage[utf8]{inputenc}
\usepackage{tabularx}
\usepackage[flushleft,alwaysadjust]{paralist}
\usepackage{enumitem}
\usepackage{etoolbox}
\usepackage{anyfontsize}
\usepackage{algorithm}
\usepackage{algpseudocode}
\usepackage{amsmath}
\usepackage{multicol}
\usepackage{graphicx}
\usepackage{tabularx}

\usepackage{tikz}

\usepackage{booktabs}
\usepackage{geometry}

\newcommand*\circled[1]{\tikz[baseline=(char.base)]{
            \node[shape=circle,draw,inner sep=1pt] (char) {#1};}}

\AtBeginDocument{%
	}

\setcopyright{acmlicensed}
\copyrightyear{2025}
\acmYear{2025}
\acmDOI{XXXXXXX.XXXXXXX}

\acmConference[SIGMOD '25]{}{June 22--27, 2025}{Berlin, DE}
\acmISBN{978-1-4503-XXXX-X/18/06}

\begin{document}
	
	\title{ARCAS: Adaptive Runtime System for Chiplet-Aware Scheduling}

	\author{Alessandro Fogli}
	\affiliation{%
		\institution{Imperial College London}
		\city{London}
		\country{United Kingdom}
	}
	\email{a.fogli18@imperial.ac.uk}
	
	\author{Bo Zhao}
	\affiliation{%
		\institution{Aalto University}
		\city{Espoo}
		\country{Finland}
	}
	\email{bo.zhao@aalto.fi}
	
	\author{Peter Pietzuch}
	\affiliation{%
		\institution{Imperial College London}
		\city{London}
		\country{United Kingdom}
	}
	\email{prp@imperial.ac.uk}
	
	\author{Jana Giceva}
	\affiliation{%
		\institution{TU Munich}
		\city{Munich}
		\country{Germany}
	}
	\email{jana.giceva@in.tum.de}
	
	\renewcommand{\shortauthors}{Fogli et al.}
	\renewcommand\abstractname{\MakeUppercase{Abstract}}

\begin{abstract}
  The growing disparity between CPU core counts and available memory bandwidth has intensified memory contention in servers. This particularly affects highly parallelizable applications, which must achieve efficient cache utilization to maintain performance as CPU core counts grow. Optimizing cache utilization, however, is complex for recent chiplet-based CPUs, whose partitioned L3 caches lead to varying latencies and bandwidths, even within a single NUMA domain. Classical NUMA optimizations and task scheduling approaches unfortunately fail to address the performance issues of chiplet-based CPUs.
		
  We describe Adaptive Runtime System for Chiplet-Aware Scheduling (ARCAS), a new runtime system designed for chiplet-based CPUs. ARCAS combines chiplet-aware task scheduling heuristics, hard\-ware-aware memory allocation, and fine-grained performance monitoring to optimize workload execution. It implements a lightweight concurrency model that combines user-level thread features—such as individual stacks, per-task scheduling, and state management—with coroutine-like behavior, allowing tasks to suspend and resume execution at defined points while efficiently managing task migration across chiplets. Our evaluation across diverse scenarios shows ARCAS's effectiveness for optimizing the performance of memory-intensive parallel applications. 
\end{abstract}

	\maketitle
	\pagestyle{plain}

    \input{1-Introduction/Introduction}
	
	\input{2-Background/Background}

	\input{3-Scheduling/Scheduling}

	\input{4-Implementation/Implementation}

	\input{5-Experiments_and_Analysis/Experiments}

	\input{7-Related_Work/Related-work}

	\input{8-Conclusion/Conclusion}
	
	
	
	
	\bibliographystyle{ACM-Reference-Format}
	\bibliography{sample-base}


\end{document}

%% file: 1-Introduction/Introduction.tex
\begin{figure}[tb]
    \centering
    \includegraphics[width=\columnwidth]{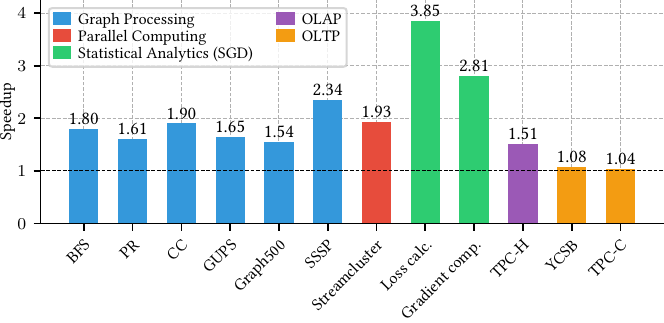}
    \caption{ARCAS speedups compared to NUMA-aware systems across various benchmarks and workloads.}
    \label{fig:introplot}
    \vspace{-0.5cm}
\end{figure}


\section{INTRODUCTION}


The rapid evolution of server hardware has led to a growing disparity between the number of CPU cores and the available memory bandwidth. This widening gap between compute power and memory resources particularly affects highly-parallelizable applications, which require efficient memory accesses. As CPU core counts continue to increase, the competition for memory bandwidth intensifies, making effective cache utilization critical for maintaining application performance \cite{Wang2016PredictingTM, elephant, Lepers2023JohnnyCT}. 

To address these challenges, researchers have explored various approaches, with a significant focus on Non-Uniform Memory Access (NUMA) architectures ~\cite{Calciu, Kashyap2017ScalableNB, Psaroudakis2016AdaptiveND}. NUMA systems attempt to bridge the gap between compute power and memory resources by providing localized memory access for each core, thus reducing contention for memory bandwidth and interconnect communication. For example, Lozi et al.~\cite{Lozi2016TheLS} provide an analysis of the Linux scheduler on NUMA systems and highlight issues with load balancing across CPU cores. Leis et al.~\cite{Leis2014MorseldrivenPA} introduce Morsel-Driven Parallelism, a NUMA-aware query evaluation framework that dynamically assigns small fragments of input data to worker threads. Kaestle et al.~\cite{Kaestle2015ShoalSA} propose Shoal, a system to automatically allocate and replicate data in NUMA systems.


While these NUMA-focused solutions have been valuable, the landscape of CPU architecture has evolved significantly with the introduction of  \emph{chiplet-based CPUs}~\cite{AMD7002, AMD7003, 2.2AMD, intelLLC, Velten2022MemoryPO}. Chiplet-based CPUs adopt a modular design that consists of multiple smaller dies, or \emph{chiplets}, interconnected to form a larger, more powerful processor. This offers superior flexibility and scalability, allowing manufacturers to create a wide range of processor configurations tailored to various performance and power requirements. Consequently, major vendors such as AMD, Intel, and ARM adopt this approach in their latest CPU designs\cite{AMD7002, AMD7003, intelLLC, Intel, ARM}.

However, chiplet-based CPUs also present new challenges in terms of memory hierarchy and inter-core communication, even within the same NUMA domain~\cite{Fogli2024OLAPOM,chipThermal,chipThermal3,intelAIB}. The partitioned nature of the L3 cache, which is distributed across multiple chiplets, introduces a trade-off between cache locality and total cache size \cite{Fogli2024OLAPOM}. Tasks scheduled on fewer chiplets benefit from lower latency due to local cache access but have limited cache capacity. Conversely, spreading tasks across multiple chiplets increases available cache size at the cost of higher access latencies. Such heterogeneity extends to inter-core communication, with a wide range of latencies and bandwidths between cores depending on their location within the chiplet architecture.
As a result, poor task scheduling policy can lead to poor performance or even degradation with increasing core counts, which is particularly acute for memory-intensive workload. Optimizing cache usage and memory allocation for heterogeneous chiplet-based architectures must therefore consider the unique hardware features and  dynamically adapt to changing workload characteristics. 

To address these challenges, we propose \emph{Adaptive Runtime System for Chiplet-Aware Scheduling}~(ARCAS), a new runtime system designed specifically for chiplet-based CPUs. ARCAS incorporates chiplet-aware task scheduling heuristics, adaptive cache partitioning, coroutine-based task management and fine-grained performance monitoring to optimize workload execution on modern heterogeneous processors:



    
    

    

\myparr{(1)~Chiplet-aware task scheduling heuristics:} ARCAS employs a heuristics that optimizes task placement based on cache affinity and inter-chiplet latencies with focus on minimizing inter-chiplet communication overhead.

\myparr{(2)~Adaptive cache partitioning:} ARCAS dynamically adjust cache allocations based on workload characteristics. This helps to balance cache locality and utilization in response to changing workload demands.

\myparr{(3)~Fine-grained parallelism:} ARCAS utilizes lightweight coroutines that offer a low context switch overhead. This enables efficient management of high concurrency, which is particularly beneficial for chiplet architectures where minimizing the overhead of task switching is crucial.

\myparr{(4)~Performance profiling and optimization:} ARCAS continuously monitors and analyzes application performance metrics. With low-overhead instrumentation, ARCAS collects data on computational load and communication patterns, which is used to inform dynamic runtime decisions (\eg task migration and scheduling adjustments).

\tinyskip
We evaluate ARCAS across diverse computational scenarios (\eg workloads from graph processing, statistical analytics, analytical databases, and high parallel processing). Based thereon, we provide insights on how to design and configure future systems to best exploit chiplet-based CPUs:

\mypar{1} Chiplet-aware task partitioning is particularly effective for workloads with irregular memory access patterns, such as graph processing algorithms.


\mypar{2} For OLAP workloads, a hybrid cache partitioning strategy often outperforms strictly local approaches.



\mypar{3} Higher core counts amplify synchronization and inter-chiplet communication overheads and therefore, affect performance. This is particularly noticeable in statistical analytics for machine learning tasks.

\mypar{4} Overly strict NUMA-aware optimizations can significantly harm performance on chiplet-based CPUs.

\tinyskip

\noindent The rest of the paper is structured as follows:

\begin{compactitem}
\item We give background on chiplet-based CPU architectures and analyze their inter-core latencies ~(\S\ref{sec:background}).
\item We present the challenges of implementing effective chiplet-aware scheduling strategies and what is relevant~(\S\ref{sec:scheduling}).
\item We detail the architecture of ARCAS, including its core components, chiplet-aware task scheduling heuristics and monitoring strategies~(\S\ref{sec:implementation}).
\item We present a comprehensive evaluation of ARCAS across various benchmarks and applications ~(\S\ref{sec:eval}).
\end{compactitem}

%% file: 2-Background/Background.tex
\section{CHIPLET-BASED CPUS}
\label{sec:background}

\begin{figure}[t]
    \centering
    \includegraphics[width=0.9\columnwidth]{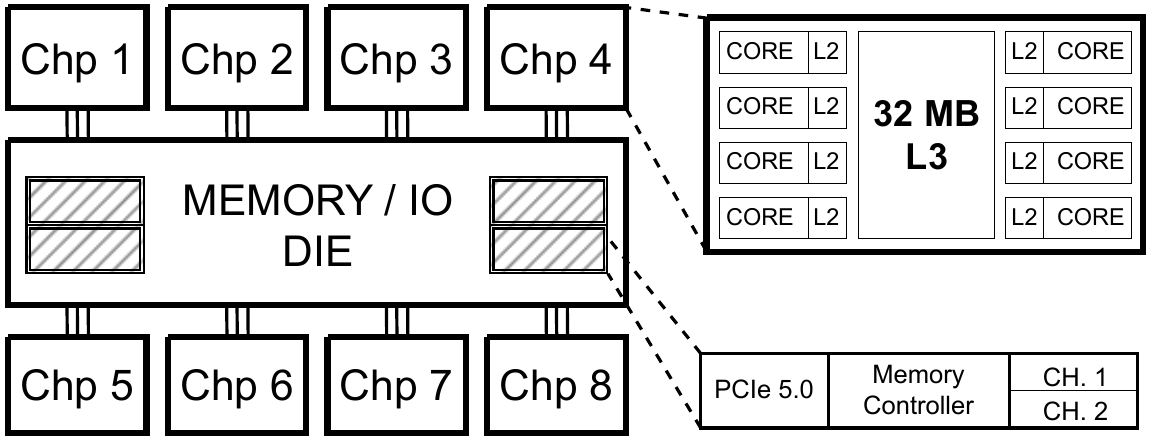}
    \caption{AMD EPYC Milan.}
    \label{fig:milan}
    \vspace{-0.5cm}
\end{figure}



Chiplet-based CPUs represent a significant evolution in processor design. 
These are modular systems that consist of multiple smaller dies, or \emph{chiplets}, that are interconnected to form a larger, more powerful processor. This design offers several advantages over traditional monolithic designs, including increased scalability, improved yield, and reduced cost \cite{Fogli2024OLAPOM}. One of the main benefits of chiplet-based designs is their flexibility. Manufacturers can combine different chiplets to create a variety of processor configurations, meeting diverse performance and power needs \cite{chipThermal}. This modular design also simplifies upgrades and modifications since individual chiplets can be replaced or updated without overhauling the entire processor. For those reasons, in recent years, several major companies like AMD, Intel, and ARM adopted this technology in their latest processors \cite{AMD7003, Intel, ARM}. 

\F\ref{fig:milan} shows the architecture of a single-socket AMD EPYC Milan CPU. It uses multiple Core Complex Dies (CCDs) (i.e., chiplets) connected to a central I/O die via AMD's Infinity Fabric interconnect. Each CCD contains one or two Core Complexes (CCXs), depending on the processor generation. Each CCX consists of multiple cores, each paired with its own L2 cache, surrounded by a shared 32 MB L3 cache. The aggregate L3 cache is distributed across the chiplets, rather than being a single unified cache for the entire processor.

\subsection{Inter-core latencies}
\label{sec:chiplet-cpus}

\begin{figure}[t]
    \centering
    \includegraphics[width=\columnwidth]{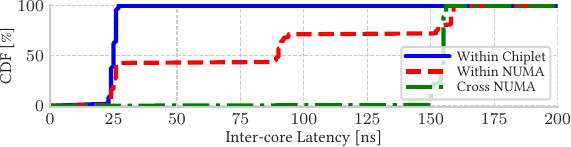}
    \caption{Cumulative Distribution Function (CDF) of core-to-core latency in an AMD EPYC Milan CPU.}
    \label{fig:cdf}
    \vspace{-0.3cm}
\end{figure}

The performance of chiplet-based CPUs is significantly influenced by the varying access times to L3 caches and the differences in latencies and bandwidths between cores. \F\ref{fig:cdf} shows the cumulative distribution function (CDF) of inter-core latencies for different communication scenarios within a  dual-socket AMD EPYC Milan CPU.  Results reveal a latency hierarchy: "Within Chiplet" communication is the fastest, followed by "Within NUMA," with "Cross NUMA" showing the highest latencies, as on-chip communication is typically faster than inter-chip communication. 

In contrast to common assumptions, the "Within NUMA" curve shows greater variability. There are three clear groupings of latencies: the lowest around 25\unit{ns}  representing intra-chiplet communication, a middle group around 80-90\unit{ns} indicating inter-chiplet but intra-CCX (Core Complex) communication, and a higher group beyond 150\unit{ns} representing communication across different CCXs within the same NUMA node. This stepped distribution highlights the heterogeneous nature of core-to-core latencies within a single NUMA domain in chiplet-based CPUs and can have a significant impact on the performance of the systems. It complicates the task allocation and resource assignment process, and it can lead to imbalances in resource utilization. 



\subsection{More cores, limited memory channels}


By utilizing chiplets, manufacturers can pack more cores into a single package, enhancing computational power and efficiency. However, this strategy does not help with the limited memory bandwidth per core. This is partly due to the limited number of memory channels available to feed these increasingly dense core configurations. 

\begin{figure}[t]
    \centering
    \includegraphics[width=0.9\columnwidth]{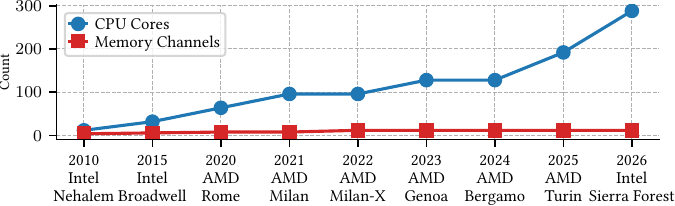}
    \caption{Number of memory channels vs. cores over the years.}
    \label{fig:dimmvscores}
    \vspace{-0.3cm}
\end{figure}

\F\ref{fig:dimmvscores} shows this increasing disparity. While back in 2010, high-end server processors typically had 4-8 cores, as of 2023, AMD's EPYC Genoa processors offer up to 96 cores, while Intel's Xeon Scalable processors provide up to 64 cores. In contrast, memory channel growth has not kept pace ---current high-end CPUs support up to 8-12 memory channels. This trend is expected to continue: by 2026, we may see CPUs with 300 cores but no more memory channels.

Looking at the cache hierarchy, particularly at the L3 level, we see that chiplet designs have enabled significant increases in cache capacity. For instance, AMD's EPYC processors feature up to 256MB of L3 cache. This helps mitigate some of the memory bandwidth limitations by keeping more data closer to the cores, but it comes with the increased heterogeneity of latencies. This needs us to adjust how we run applications to unlock the full potential of chiplet-based CPUs.

\subsection{Parallel processing on chiplets} 
\label{sec:working_set_chiplets}

The increasing core count and the limited memory bandwidth present challenges for efficient parallel processing on chiplet-based CPUs. As more cores compete for the same memory resources, intelligent utilization of on-chip caches becomes crucial to maintain high performance. This scenario raises important questions about how to best leverage the distributed nature of L3 caches across multiple chiplets.

We investigate the performance impact of two caching strategies on multi-chiplet CPU architectures: \texttt{LocalCache} and \texttt{Distributed-Cache}. In our setup, \texttt{LocalCache} confines data access to CPU cores within a single chiplet, thereby leveraging its local L3 cache. \texttt{DistributedCache} distributes the same number of cores across multiple chiplets, thus utilizing the collective L3 cache capacity, but incurring inter-chiplet communication overhead.
To understand the effects, we conducted a microbenchmark on a single-socket AMD Epyc Milan processor. The benchmark measured the execution time of a multithreaded write operation on a vector, divided into chunks processed by 8 cores across 1,000 iterations, varying the data size from 38\unit{B} to 38\unit{GB}.

\begin{figure}[t]
    \centering
    \includegraphics[width=\columnwidth]{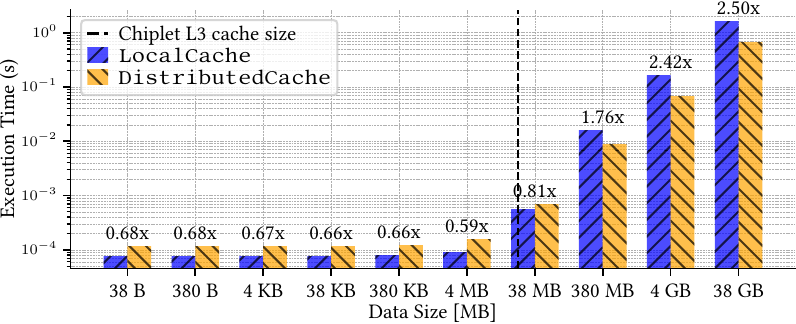}
    \caption{\texttt{LocalCache} vs. \texttt{DistributedCache}: write operation speedup when varying the data array size.}
    \label{fig:similSTREAM}
    \vspace{-0.5cm}
\end{figure}

Results, displayed on a logarithmic scale in \F\ref{fig:similSTREAM}, indicate that for data sizes up to 38 MB, \texttt{LocalCache} outperforms \texttt{DistributedCache} with lower execution times due to the avoidance of inter-chiplet communication. This advantage diminishes beyond 32 MB, when we exceed the L3 cache capacity. At that point, the \texttt{DistributedCache} becomes more effective, peaking at a 2.50$\times$ speedup for the 38\unit{GB} dataset. This shift underscores the trade-off between leveraging local cache access and the expanded capacity offered by modern chiplet-based processors, that the range in performance between the two strategies is between 0.59$\times$ and 2.50$\times$, and that which one is better highly depends on the working set size and the access patterns of the workload.

%% file: 3-Scheduling/Scheduling.tex
\section{DESIGNING CHIPLET-AWARE RUNTIME SYSTEMS}
\label{sec:scheduling}

Runtime systems are an effective solution to the challenges of parallel computing in modern hardware environments. At their core, these systems decompose applications into discrete units of work, or tasks, with well-defined inputs and outputs. Rather than directly calling computation kernels, developers define tasks and their dependencies, allowing the runtime system to dynamically distribute work across the available compute resources. 

The strength of task-based systems lies in their dynamic approach to scheduling. Advanced runtime systems, integrate detailed information about task memory footprints, dependencies, and execution characteristics. This allows for more informed scheduling decisions, reducing scheduling overhead and better management of memory hierarchies. Such optimization is especially relevant for high-parallel tasks, where inefficient scheduling can lead to performance degradation due to load imbalance and non-uniform memory access (NUMA) effects.

In addition, task-based runtime systems can significantly improve the programmability and portability of parallel applications. By abstracting away the complexities of task scheduling and hardware-specific optimizations, developers can focus on expressing parallelism in their applications without concern for the intricacies of the underlying hardware. This separation of concerns is important for developing scalable and maintainable parallel software in an era of rapidly evolving hardware architectures.

\subsection{Challenges with chiplets}
\label{challenges_charm}

A chiplet-aware runtime system needs to address a number of challenges: 

\tinyskip
\noindent\myparr{(1). Cache management:} Unlike memory, users cannot directly allocate data in chiplet caches. Traditional runtime systems have primarily focused on memory allocation in NUMA nodes or thread placement based on core availability and workload distribution, but chiplet caches require a different approach. Existing cache management schemes are insufficient for chiplet architectures as they fail to adequately consider the trade-off between cache locality and cache availability, especially when the cache is partitioned (\S\ref{sec:background}). Technologies like Intel's CAT are used to partition the cache for better isolation between concurrent processes, rather than data allocation \cite{intelCAT}.

\tinyskip
\noindent\myparr{(2). Inter-chiplet communication and synchronization:} The increased latency between chiplets showed in \S\ref{sec:chiplet-cpus} requires optimized task mapping and scheduling. Runtime systems need to minimize data movement across chiplets, but also implement efficient and lightweight synchronization mechanisms to prevent bottlenecks. Traditional thread-based methods are inadequate for chiplet architectures due to the overhead of OS threads in creation, context switching and management. This is especially problematic for fine-grained tasks when the number of chiplets grows. 

\tinyskip
\noindent\myparr{(3). Workload adaptivity:} A single policy cannot address the diverse needs of all applications. A workload's requirements can also change dynamically. Some are latency-sensitive, requiring fast data access. Others are memory-intensive, demanding large amounts of fast memory accesses.  Moreover, the working set size can fluctuate during execution. This variability necessitates adaptive resource management (see \S\ref{sec:working_set_chiplets}). For instance, a workload might initially benefit from high data locality, keeping related data close. However, as the working set expands, the same workload may require increased cache availability to maintain performance. Effective chiplet-aware runtime systems need to be capable of recognizing these shifting demands and adjust the resource allocation accordingly throughout the execution lifecycle.

%% file: 4-Implementation/Implementation.tex
\section{ARCAS SYSTEM DESIGN}
\label{sec:implementation}


We propose ARCAS, a runtime system to tackle the issues of cache management, inter-chiplet communication, and workload adaptivity in modern chiplet-based CPUs.

At its core, ARCAS revolves around two fundamental pillars: (1) \textbf{Lightweight task management}: ARCAS implements nimble mechanisms for task placement and synchronization (see \S\ref{scheduling} and \S\ref{adaptive}), seamlessly integrating with contemporary runtime systems. This approach enables granular control over task execution across chiplets, minimizing overhead and maximizing efficiency.  (2) \textbf{Intelligent placement decisions}: By incorporating chiplet-aware objective evaluation criteria, ARCAS makes informed choices about task movement (see \S\ref{coroutines}), improving utilization of chiplet-specific interconnects and overall system performance.

To handle the variability in workload requirements, ARCAS adopts an adaptive approach. It incorporates a lightweight profiling component that gathers key metrics about task behavior, including memory access patterns and communication frequency. Based on this profiling data, ARCAS dynamically adjusts its task placement strategies at runtime. This enables ARCAS to adapt to workload changes and system conditions, ensuring efficient resource use across evolving computational demands.

\begin{figure}[t]
    \centering
    \includegraphics[width=0.95\columnwidth]{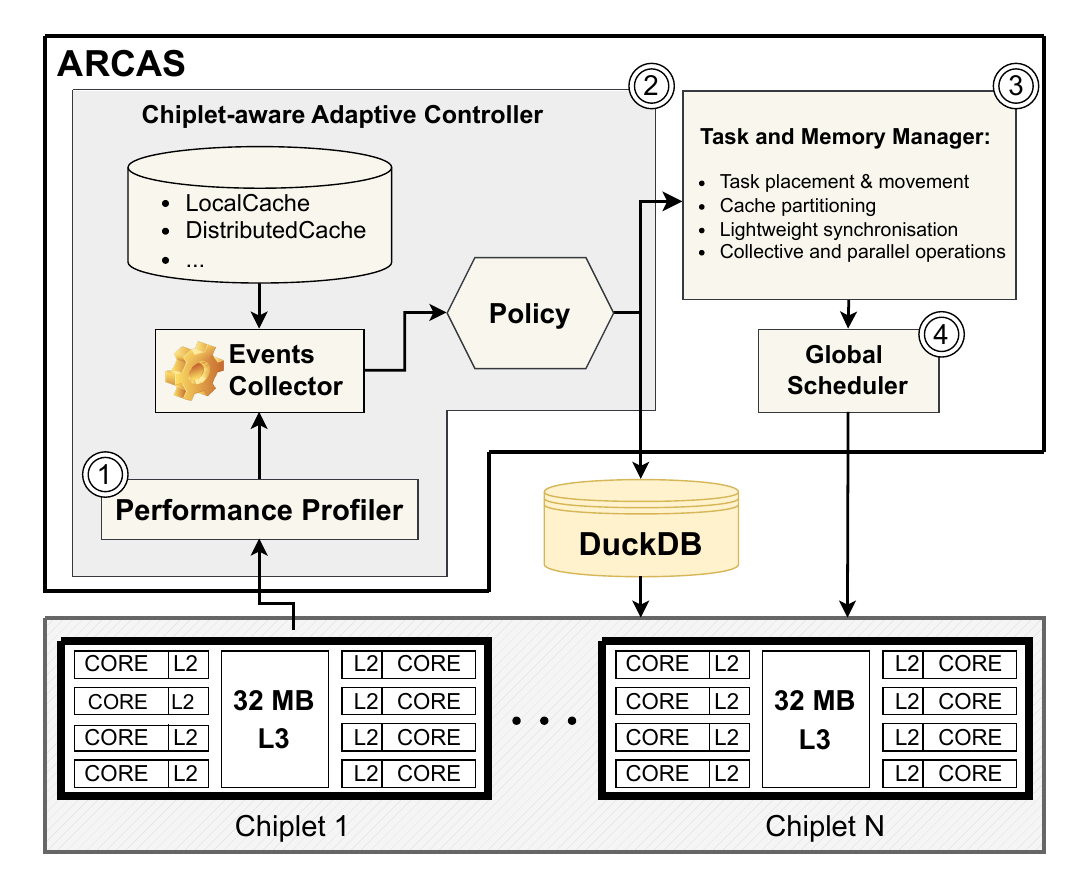}
    \caption{ARCAS Architecture.}
    \label{fig:charm}
    \vspace{-0.5cm}
\end{figure}
 
\subsection{ARCAS architecture}




The ARCAS system is designed with four key components. As illustrated in \F\ref{fig:charm}, the architecture includes: (1) a performance profiler, (2) an adaptive controller, (3) a memory and task manager, and (4) a global scheduler.

The \textbf{performance profiler} \circled{1} continuously monitors and analyzes application performance metrics. Using low-overhead instrumentation, ARCAS collects detailed data on computational load and communication patterns. During workload execution, we periodically check the frequency of accesses to the local chiplet, remote chiplets, and main memory. This helps us determine the cache availability. The profiler can also monitor only specific code segments, providing detailed and accurate results for individual tasks or threads with minimal overhead.

The \textbf{adaptive controller} \circled{2} gathers information from the profiler and uses predefined approaches to generate scheduling policies. An approach outlines the general method or guiding principle, while a policy specifies the concrete actions the scheduler follows based on that approach. For example, a location-centric approach might focus on minimizing cross-chiplet communication, while the corresponding policy could dictate assigning tasks to cores within the same chiplet. These predefined approaches can be extended to create more precise policies tailored to workflow requirements. In our runtime system, the controller generates adaptive policies that switch between location-centric and cache size-centric approaches. These policies dynamically balance the benefits of local cache access with the need for larger aggregate cache sizes, adapting to changing workload requirements. This adaptive partitioning approach allows ARCAS to respond to evolving workload demands, ensuring high performance and efficiency.

The \textbf{task and memory manager} \circled{3} in ARCAS efficiently handles tasks from the user's application using lightweight coroutines. These coroutines are allocated or shifted to specific CPU cores, enabling the management of numerous concurrent tasks and enhancing parallel execution and overall application throughput. The system supports NUMA-aware allocation to optimize main memory access patterns. Barrier synchronization mechanisms are also provided to coordinate task execution across multiple chiplets, ensuring synchronized task execution.

Lastly, the \textbf{global scheduler} \circled{4} coordinates task distribution and load balancing across all available compute resources. It receives policies from the adaptive controller and interfaces with the task and memory manager to obtain lightweight coroutines representing individual tasks. It also maintains a global view of the system and can quickly respond to changes in workload patterns receiving new policies and migrating tasks between chiplets as needed, ensuring the system remains optimized under varying conditions.




\subsection{Chiplet-aware task scheduling policy}
\label{scheduling}

Maximizing cache utilization is important for achieving high performance on chiplet-based systems with complex memory hierarchies. ARCAS incorporates specific heuristics to improve cache affinity when scheduling and migrating tasks. 
These heuristics aim to keep related data and computations co-located to minimize expensive data movement. To handle more dynamic workloads, ARCAS also implements an adaptive cache partitioning scheme. This monitors the cache usage of different task types and adjusts the allocation of cache resources to maximize overall system throughput. The partitioning is periodically refined based on runtime performance measurements.

The \texttt{Chiplet Scheduling Policy} (Algorithm \ref{alg:scheduling}) implements an adaptive approach to manage task distribution across chiplets based on cache usage patterns. This algorithm operates as follows: it periodically checks the time elapsed since the last scheduling decision and ensures that task reassignment occurs at regular intervals to adapt to changing workloads. If enough time has passed, the algorithm retrieves the cache fill event counter, which tracks the number of remote memory accesses between chiplets. This is important because inter-chiplet communication tends to be more expensive than intra-chiplet communication, unlike in monolithic CPUs where these costs are more uniform.

Next, the algorithm calculates the rate of these remote accesses and compares it to a predefined threshold. This threshold can vary depending on the approach used — for instance, a higher value would delay changes to the scheduling. If the rate exceeds this threshold, it indicates that the system is experiencing high inter-chiplet communication. To mitigate this, the algorithm increases the \texttt{spread\_rate} — which means distributing tasks across more chiplets. However, if the remote access rate is below the threshold, the algorithm reduces the \texttt{spread\_rate}. In this case, it compacts tasks onto fewer chiplets that can enhance cache locality by reducing the distance and frequency of memory accesses across chiplet boundaries. 

\begin{algorithm}[t]
\caption{Chiplet Scheduling Policy}
\label{alg:scheduling}
\footnotesize
\begin{algorithmic}[1]
\Procedure{ChipletScheduling}{}
    \State $current\_time \gets \text{steady\_clock.now()}$ \label{alg:time_check}
    \State $elapsed \gets current\_time - time$
    \If{$elapsed \geq \text{SCHEDULER\_TIMER}$}
        \State $counter \gets \text{getEventCounter()}$ \Comment{Cache fill events} \label{alg:counter}
        \State $rate \gets counter \times \text{SCHEDULER\_TIMER} / elapsed$ \label{alg:rate}
        \If{$rate \geq \text{RMT\_CHIP\_ACCESS\_RATE}$} \label{alg:rate_check}
            \If{$spread\_rate < \text{CHIPLETS}$}
                \State $spread\_rate \gets spread\_rate + 1$ \label{alg:spread_inc}
            \EndIf
        \Else \label{alg:rate_below}
            \If{$spread\_rate > 1$}
                \State $spread\_rate \gets spread\_rate - 1$ \label{alg:spread_dec}
            \EndIf
        \EndIf
        \State $\text{updateLocation()}$ \Comment{Spread or compact on chiplets} \label{alg:update}
        \State $time \gets \text{steady\_clock.now()}$
        \State $\text{resetEventCounter()}$
    \EndIf
\EndProcedure
\end{algorithmic}
\end{algorithm}

\begin{algorithm}[t]
\caption{Update Location}
\label{alg:update_location}
\footnotesize
\begin{algorithmic}[1]
\Procedure{UpdateLocation}{}
    \If{$\text{spread\_rate} \notin (0, \text{CHIPLETS}]$ \textbf{or} $\text{THREAD\_SIZE} \geq \text{spread\_rate} \times \text{CORES\_PER\_CHIPLET}$} \label{alg:bounds_check_start}
        \State \Return \Comment{Bounds check} \label{alg:bounds_check_end}
    \EndIf
    \State $\text{chiplet} \gets \left\lfloor \frac{\text{rank}}{\text{CORES\_PER\_CHIPLET} / \text{spread\_rate}} \right\rfloor$ \label{alg:chiplet_calc}
    \State $\text{slot} \gets \text{rank} \bmod \left(\frac{\text{CORES\_PER\_CHIPLET}}{\text{spread\_rate}}\right)$ \label{alg:slot_calc}
    \If{$\text{chiplet} \geq \text{CHIPLETS}$} \label{alg:chiplet_wrap_start}
        \State $\text{chiplet} \gets \text{chiplet} \bmod \text{CHIPLETS}$
        \State $\text{slot} \gets \text{slot} + \left\lfloor \frac{\text{rank}}{\text{CORES\_PER\_CHIPLET}} \right\rfloor$ \label{alg:chiplet_wrap_end}
    \EndIf
    \State $\text{core} \gets \text{chiplet} \times \text{CHIPLETS} + \text{slot}$ \label{alg:affinity}
    \State $\text{set\_thread\_affinity}(\text{core})$ \Comment{Set affinity}
    \State $\text{numa\_node} \gets \left\lfloor \frac{\text{core}}{\text{CORES\_PER\_NUMA\_NODE}} \right\rfloor$ \label{alg:mempolicy}
    \State $\text{set\_mempolicy}(\text{MPOL\_BIND}, 1 \ll \text{numa\_node})$ \Comment{Set memory policy}
\EndProcedure
\end{algorithmic}
\end{algorithm} 

Finally, the algorithm invokes the \texttt{update-Location()} function, which redistributes tasks according to the new \texttt{spread\_rate}. This dynamic adjustment helps to balance the trade-off between minimizing inter-chiplet communication and maximizing cache locality.

\subsection{Adaptive cache partitioning}
\label{adaptive}

ARCAS's adaptive approach allows the scheduler to balance between two competing goals: reducing inter-chiplet communication when it becomes excessive, and consolidating tasks to improve cache locality when communication is low. The \texttt{Update Location} algorithm (Algorithm 2) is responsible for implementing the task distribution strategy determined by the Chiplet Scheduling algorithm. It operates as follows: the algorithm begins by checking that the \texttt{spread\_rate} is valid and that there are enough cores to handle all threads. Next, it calculates the target chiplet and core slot for each thread based on its rank and the \texttt{spread\_rate}. If the chiplet exceeds the total amount available, it wraps around, ensuring an even distribution across chiplets.

The algorithm then sets the thread's affinity to a specific core within the chiplet, optimizing task placement to reduce inter-chiplet communication. Finally, it binds memory allocation to the correspondingg NUMA node, further minimizing remote memory accesses across chiplets.

The combination of the two presented algorithms allows ARCAS to dynamically adjust its task distribution strategy based on observed cache usage patterns. This adaptive approach can potentially improve performance for a wide range of workloads by finding an appropriate balance between spreading tasks for reduced contention and consolidating them for improved locality.





\subsection{Fine-grained task parallelism}
\label{coroutines}

Traditional OS-managed threads can introduce high overhead when context switching and synchronization, which is particularly highlighted when managing a high volume of tasks across multiple chiplets.

To overcome these challenges, ARCAS implements a lightweight concurrency system that combines features of user-level threads and coroutines. Like user-level threads, ARCAS tasks have their own scheduler, individual stacks for each execution unit, and complex state management. This design allows ARCAS to handle task movement across chiplets efficiently, managing execution contexts independently of the OS. At the same time, ARCAS incorporates coroutine-like behavior, particularly the ability to suspend and resume execution at developer-defined points, similar to the concurrency model used in the RING runtime system \cite{Meng2017RINGNM}. For instance, when a coroutine yields, ARCAS's integrated profiling system activates, analyzing task behavior, memory access patterns, and inter-chiplet communication. This profiling data enables ARCAS to dynamically adjust task placement across chiplets, optimizing cache locality and minimizing remote memory accesses in real time.

Within each core, ARCAS maintains a local task queue designed for low-overhead operation. Using lock-free mechanisms based on atomic operations, tasks are enqueued and dequeued efficiently by multiple worker threads without locks, avoiding costly synchronization delays. When a coroutine yields or completes, the worker thread checks its local deque for pending tasks. If the worker's local task queue is empty, it uses a work-stealing approach—first attempting to steal tasks from cores on the same chiplet before reaching out to other chiplets. This strategy helps preserve cache locality, which is critical for performant chiplet-based systems.

\subsection{Performance profiling and optimization}

Performance monitoring in ARCAS is designed to provide real-time insights into the system's performance, enabling dynamic adjustments to the scheduling and memory management policies. ARCAS provides a set of performance counters that can be used to monitor various aspects of the system's performance, such as cache misses, memory bandwidth utilization, and task execution times. These counters are integrated with the global scheduler, allowing for real-time performance monitoring and adaptive scheduling based on the current performance metrics.  

ARCAS's runtime system efficiently collects performance data in user space, minimizing context switches and overhead. This data drives dynamic adjustments to scheduling and memory management policies, adapting to changing workloads and computational needs. The system's effectiveness varies with workload characteristics and hardware configurations, potentially requiring tuning of thresholds and adjustment rates based on empirical observations. For instance, data-sharing-intensive tasks may benefit from consolidation, while independent tasks might perform better with distributed execution.


\subsection{Implementation and API}

ARCAS is implemented as a \textit{C++} software framework to enable efficient mapping and execution of applications on heterogeneous chiplet-based CPUs. Based on the work done on Grappa \cite{Nelson2014GrappaA} and RING \cite{Meng2017RINGNM}, the design is highly modular, with separate components handling different aspects of the system such as task scheduling, memory management, communication and profiling. Each component can be reused, reconfigured or replaced independently, allowing for a high degree of customization and optimization tailored to specific application requirements. 


ARCAS provides developers with a straightforward and efficient API for parallel programming on chiplet-based architectures. The framework is initialized using \texttt{ARCAS\_Init()} and cleaned up with \texttt{ARCAS\_Finalize()}. Developers can define parallel tasks using lambda functions within the \texttt{run()} function. ARCAS offers various task execution methods, including \texttt{all\_do()} for executing a task on all cores, and \texttt{call()} for remote procedure calls with both synchronous and asynchronous options. When executing these tasks, developers can focus on application logic while ARCAS handles the complexities of chiplet-aware resource allocation, task distribution, and performance optimization. ARCAS also provides synchronization primitives like \texttt{barrier()} to ensure that parallel tasks are coordinated effectively,  preventing unnecessary overhead.

To collect and analyze performance metrics, the ARCAS system uses the \textit{libpfm} library \cite{perfmon2}. Before starting our evaluation of ARCAS, we performed a sensitivity analysis systematically varying the candidate threshold value and measuring its impact on key performance metrics like throughput, latency and cache utilization. Based on this comprehensive analysis, we determined that an \texttt{RMT\_CHIP\_ACCESS\_RATE} of 300 events per \texttt{SCHEDULER\_TIMER} interval provided the best balance of performance across our test scenarios. This threshold effectively triggers task redistribution when inter-chiplet communication becomes excessive, while also allowing sufficient task consolidation to benefit from cache locality.

%% file: 5-Experiments_and_Analysis/Experiments.tex
\begin{figure*}[!tb]
	\centering
	\begin{tabular}{l|c|r}
		\begin{subfigure}[t]{0.315\textwidth}
			\centering
			\includegraphics[width=\linewidth]{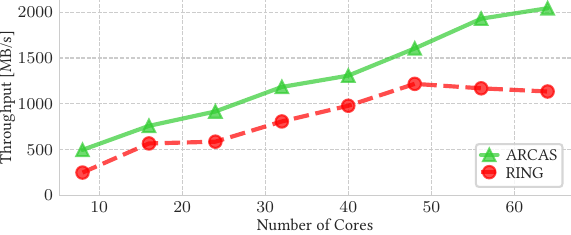} 
			\caption{Breadth-First Search (BFS)}
			\label{fig:bfsscala}
		\end{subfigure}
		&
		\begin{subfigure}[t]{0.315\textwidth}
			\centering
			\includegraphics[width=\linewidth]{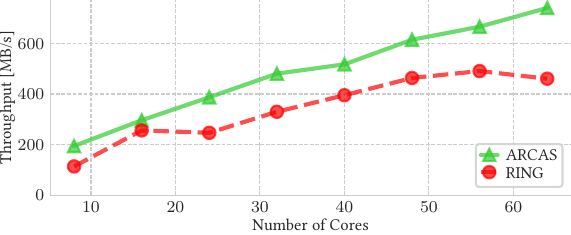}
			\caption{PageRank (PR)}
			\label{fig:prscala}
		\end{subfigure}
		&
		\begin{subfigure}[t]{0.315\textwidth}
			\centering
			\includegraphics[width=\linewidth]{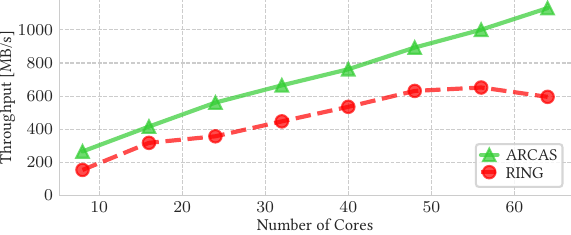}
			\caption{Connected Component (CC)}
			\label{fig:ccscala}
		\end{subfigure}
	\end{tabular}
	\begin{tabular}{l|c|r}
		\begin{subfigure}[t]{0.315\textwidth}
			\centering
			\includegraphics[width=\linewidth]{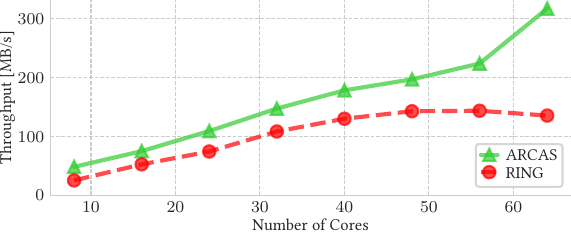} 
			\caption{Single Source Shortest Path (SSSP)}
			\label{fig:ssspscala}
		\end{subfigure}
		&
		\begin{subfigure}[t]{0.315\textwidth}
			\centering
			\includegraphics[width=\linewidth]{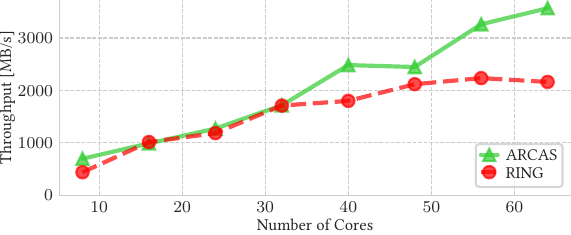}
			\caption{RandomAccess (GUPS)}
			\label{fig:gupsscala}
		\end{subfigure}
		&
		\begin{subfigure}[t]{0.315\textwidth}
			\centering
			\includegraphics[width=\linewidth]{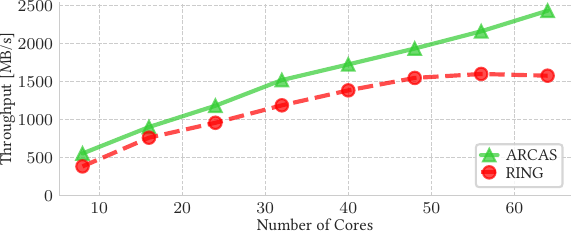}
			\caption{Graph500}
			\label{fig:graph500scala}
		\end{subfigure}
	\end{tabular}
	\caption{Graph Processing + Random Access Scalability}
	\label{fig:graphscalability}
\end{figure*}

\section{EXPERIMENTAL ANALYSIS}
\label{sec:eval}

We have conducted a comprehensive evaluation of ARCAS's efficacy in modern chiplet-based architectures over a range of benchmarks and applications. 
In particular, we address the following questions:

\tinyskip
\begin{compactenum}
	\item[\textbf{Q1:}] How does ARCAS's performance compare to existing runtime systems including RING~\cite{Meng2017RINGNM} and Shoal~\cite{Kaestle2015ShoalSA} across different computational tasks? (\S\ref{sec:eval:Graph})
	
		\item[\textbf{Q2:}] What are the performance trade-offs and scalability of ARCAS on chiplet-based architectures compared to optimized NUMA-aware systems \eg Shoal~\cite{Kaestle2015ShoalSA}? (\S\ref{sec:eval:Shoal})
	
	\item[\textbf{Q3:}] How effectively can ARCAS exploit the heterogeneous chiplet-based architectures to accelerate applications with irregular computation and memory access patterns, \eg graph processing and sparse linear algebra? (\S\ref{sec:eval:SGD})
	
		\item[\textbf{Q4:}] How does ARCAS accelerate data-intensive analytical workloads (\S\ref{sec:eval:OLAP}) and does it help with transactional workloads  (\S\ref{sec:eval:OLTP})?
		
		
	
\end{compactenum}

\subsection{Experimental setup}
\label{sec:expSetup}
\mypar{Testbed} We have executed the experiments on a dual-socket AMD EPYC Milan 7713 processor. Each socket features 64 CPU cores, 512 \unit{GB} RAM and 8 chiplets --- each chiplet is equipped with a 32\unit{MB} local L3 cache.  The codebase is compiled in GCC 12 with the \texttt{-O3} optimization flag, running in Ubuntu 23.04.

\mypar{Baselines} We compare the performance of ARCAS against two baseline systems:
\begin{compactenum}
	\item RING~\cite{Meng2017RINGNM} is a NUMA-aware, message-batching runtime system designed for high-performance and in-memory data-intensive workloads.
	\item SHOAL~\cite{Kaestle2015ShoalSA} is a runtime system that provides an array abstraction for optimized memory allocation and access patterns on NUMA multi-core architectures.
\end{compactenum}

\mypar{Benchmarks} To evaluate ARCAS and baseline systems, we use workloads from four key areas, (i) graph processing, (ii) high performance parallel processing, (iii) statistical analytics and (iv) database management systems. Specifically, we use the following benchmarks and applications: 

\begin{compactitem}
	\item \emph{RandomAccess} evaluates the performance of non-contiguous memory access in a distributed shared memory architecture, measured in global updates per second (GUPS).\looseness=-1
	
	\item \emph{Graph algorithms} assess performance on irregular access patterns. We have tested five graph algorithms including Breadth-First Search (BFS)~\cite{moore1959bfs}, Connected Component (CC)~\cite{Xie2023OnQC}, Single Source Shortest Path (SSSP)~\cite{SSSP}, PageRank (PR)~\cite{brin1998pagerank} and Graph500~\cite{murphy2010graph500}. The graph used for the benchmark is a Kronecker graph model with $2^{24}$ vertices and 16$\times2^{24}$ edges, approximately 4\unit{GB}.
	
    \item \emph{Statistical analytics} evaluate the performance of Stochastic Gradient Descent (SGD) on a dataset with 10,000 samples and 8,192 features, totaling approximately 6,250\unit{MB}. For this evaluation, we use DimmWitted, an analytics engine optimized for statistical computations on modern NUMA architectures \cite{Zhang2014DimmWittedAS}.
	
	\item \emph{OLAP workload} evaluates TPC-H~\cite{tpch2021}, with scale factor 100, using the DuckDB~\cite{Raasveldt2019DuckDBAE} engine with and without the ARCAS adaptive controller module.

 	\item \emph{OLTP workload} evaluates the YCSB~\cite{Cooper2010BenchmarkingCS} and TPC-C~\cite{tpcc2010} benchmarks using the Ermia~\cite{Kim2016ERMIAFM} engine. YCSB is configured with 50 million records in a single table, running a mixed workload of 45\% read and 55\% read-modify-write operations. TPC-C simulates 50 warehouses with a workload of 45\% New Order, 43\% Payment, and smaller proportions of Delivery, Order Status, and Stock Level transactions. It supports cross-partition transactions, uses a uniform item distribution, and always accesses the home warehouse. The implementation includes 10,000 suppliers without specialized scan optimizations for Order Status.
  
	\item The \emph{Streamcluster benchmark} \cite{Zhan2017PARSEC30AM} models compute-intensive clustering algorithms sensitive to memory access patterns. We use it to compare ARCAS with SHOAL\cite{Kaestle2015ShoalSA} in parallel processing scenarios on shared-memory multicore architectures. Streamcluster provides insights into working sets, locality, data sharing, synchronization, and off-chip traffic. The test processes 1 million data points with 128 dimensions, targeting 10–20 cluster centers and allowing up to 5,000 intermediate centers, with data chunked into 200,000-point batches.
\end{compactitem}

\mypar{Measurements} We measure three performance indicators including (i) the throughput, (ii) the speedups of ARCAS over baseline systems, and (iii) the memory bandwidth utilization. The results are obtained as the average of 10 runs.


\subsection{Overall effectiveness and efficiency}
\label{sec:eval:Graph}
We first investigate the overall effectiveness and efficiency of ARCAS.
To this end, we examine the scalability of the processing throughput of six different algorithms ---  Breadth-First Search (BFS), PageRank (PR), Connected Component (CC), Single Source Shortest Path (SSSP),  RandomAccess (GUPS), and Graph500 --- and compare ARCAS's scalability against the baseline of RING.

 \F\ref{fig:graphscalability} shows the results. We observe that ARCAS achieves near linear scalability for throughput and scales better than RING, where the margin widens for larger core counts. The most pronounced speedups for ARCAS over RING are observed in the BFS, CC, and SSSP algorithms, achieving speedups of 1.8$\times$, 1.9$\times$ and 2.3$\times$, respectively.
 
We attribute this to its advanced chiplet-aware task scheduling policy to optimize inter-chiplet communication and the load balancing strategies to exploit the local L3 cache. Specifically, while RING's NUMA-aware design avoids remote memory allocation and accesses, it fails to prevent the L3 cache access from remote NUMA domains.
This results in increased communication overhead (\ie data movement) of accessing L3 cache across multiple chiplets and sockets. 
The effects become more evident for the high core counts on all benchmarks in \F\ref{fig:graphscalability}. 

In contrast, ARCAS significantly reduces remote NUMA chiplet accesses (see \T\ref{tab:ring}). 
ARCAS collocates tasks and data into local chiplets and therefore, avoids the NUMA-negative effect. Since each NUMA is equipped with 64 cores, ARCAS fully occupies all cores in a single socket. 
This eliminates "free spots" for core/task replacement, reducing the movement of processes across chiplets in the runtime.
For instance, in the SSSP benchmark, ARCAS performed $1.5\times10^8$ accesses at local chiplets compared to RING's $8.7\times10^4$ accesses, while remote accesses are just $6.0\times10^3$ for ARCAS but $2.3\times10^8$ for RING. Such trade-off between task allocation/movement and cache locality results in ARCAS's superior scalability. \looseness=-1

\begin{table}[]
\centering
\footnotesize
\caption{Comparison of chiplet accesses ($\times 10^3$) using 64 cores.}
\resizebox{\columnwidth}{!}{%
\begin{tabular}{lrrrr}
\toprule
\textbf{Application} & \multicolumn{2}{c}{\textbf{Remote NUMA Chiplet}} & \multicolumn{2}{c}{\textbf{Local Chiplet}} \\
\cmidrule(r){2-3} \cmidrule(r){4-5}
 & \textbf{ARCAS} & \textbf{RING} & \textbf{ARCAS} & \textbf{RING} \\
\midrule
BFS     & 3      & 20876   & 24722   & 14687   \\
PR      & 9      & 55960   & 78561   & 39212   \\
CC      & 84     & 43718   & 54631   & 27924   \\
SSSP    & 6      & 230939  & 153665  & 87152   \\
GUPS    & 25     & 19377   & 21033   & 18328   \\
Graph500& 93     & 93196   & 229608  & 176853  \\
\bottomrule
\end{tabular}
} %
\label{tab:ring}
\end{table}

\begin{figure}[t]
    \centering
    \includegraphics[width=0.9\columnwidth]{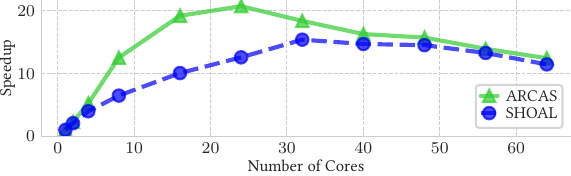}
    \caption{Scalability Analysis of StreamCluster Performance: Shoal Compared to ARCAS.}
    \label{fig:shoal}
\end{figure}

ARCAS effectively exploits  resources of chiplet-based architectures, because of the chiplet-aware optimized task scheduling policy and load balancing mechanisms. The linear scalability in iterative algorithms (\eg PageRank) manifests ARCAS's efficient synchronization schemes to reduce the overhead of global updates and allow for better parallelism across chiplets. ARCAS's performance advantage is reflected across different graph algorithms, indicating its flexibility to adapt to varying computational patterns and data dependencies. 


\subsection{Scalability comparison with optimized NUMA-aware systems}
\label{sec:eval:Shoal}

To further investigate the impact of chiplet-aware task scheduling on parallel workloads, we evaluate ARCAS over the StreamCluster benchmark~\cite{Zhan2017PARSEC30AM} and compare its performance with SHOAL~\cite{Kaestle2015ShoalSA}.
Recall that SHOAL optimizes the memory allocation and replication on NUMA architectures based on the hints of both the architecture properties (\eg huge pages and DMA copy engines) and the workload. 

 
\F\ref{fig:shoal} demonstrates the speedups of ARCAS and SHOAL, ranging the core counts from 1 to 64.  Here, ARCAS achieves the highest speedup of 21$\times$ at 24 cores, while the best speedup of SHOAL is 16$\times$ at 32 cores.  The biggest performance gap is observed at 16 cores when ARCAS's performance is 2$\times$ of SHOAL.  ARCAS maintains its advantage over SHOAL until 40 cores. Beyond that, ARCAS's performance drops slightly but  still outperforms SHOAL till 64 cores. 

We attribute these effects to the bespoke approaches of task assignment, core and cache utilization, especially in the chiplet-based architectures (\ie AMD EPYC Milan). Specifically, SHOAL employs a NUMA-aware task-to-core assignment scheme to allocate tasks sequentially to cores in numerical order (\ie task 0 to core 0, task 1 to core 1, etc). 
While such an approach indeed optimizes for  NUMA architectures, it significantly limits SHOAL's ability to exploit the heterogeneous chiplet-based architectures. For instance, when using 16 cores, SHOAL confines computation to just two chiplets even though eight chiplets are available, resulting in significant smaller cache sizes---uses $2\times32$\unit{MB} L3 cache despite $8\times32$\unit{MB} are available---and therefore,  the limited performance. In this test, we used 1 million points, each with 128 dimensions, where each dimension is a 4-byte floating-point number, resulting in a total dataset size of approximately 512\unit{MB}, which far exceeds the cache capacity of two chiplets.

In contrast, ARCAS's chiplet-aware design allows for dynamically distributing tasks across multiple chiplets. This enables ARCAS to fully exploit the entire resources of the processor, especially the distributed L3 caches across all chiplets. For instance, at 16 cores, ARCAS places the tasks across all eight chiplets of AMD EPYC Milan, leveraging $8\times32$\unit{MB} total L3 cache size (compared to SHOAL's  $2\times32$\unit{MB}) which significantly reduces the access from the main memory. This is more pronounced for smaller core counts (\eg below 40 on AMD EPYC Milan), where SHOAL's limited chiplet utilization is more evident.  Increasing core counts (\eg above 40) makes such difference less noticeable, as SHOAL gradually utilizes L3 caches from additional chiplets. To corroborate, we present a zoom-in analysis of the cache/memory access patterns in \T\ref{tab:shoal}, measuring the number of accesses to L3 caches in local chiplet, remote chiplet but within a NUMA domain, and to main memory. 
At 8 cores, SHOAL shows over $7\times$ more main memory accesses than ARCAS. This is because SHOAL uses 8 cores within 1 chiplet and frequently fetches data from the main memory.  In contrast, ARCAS places each core (\ie task) on separate chiplets (8 chiplet in total), leveraging $8\times$ L3 cache size than SHOAL. \looseness=-1

\begin{table}[t]
\centering
\footnotesize
\caption{Comparison of memory and cache accesses ($\times 10^3$) between ARCAS and SHOAL across different core counts.}
\resizebox{\columnwidth}{!}{%
\begin{tabular}{lrrrrrr}
\toprule
\textbf{Cores} & \multicolumn{2}{c}{\textbf{Local Chiplet}} & \multicolumn{2}{c}{\textbf{Local NUMA Chiplet}} & \multicolumn{2}{c}{\textbf{Main Memory}} \\
\cmidrule(r){2-3} \cmidrule(r){4-5} \cmidrule(r){6-7}
 & \textbf{ARCAS} & \textbf{SHOAL} & \textbf{ARCAS} & \textbf{SHOAL} & \textbf{ARCAS} & \textbf{SHOAL} \\
\midrule
8   & 27055  & 11014  & 718    & 9      & 7037   & 49222  \\
16  & 30956  & 27302  & 893    & 3623   & 6941   & 41273  \\
32  & 35099  & 50054  & 1179   & 2528   & 6634   & 24832  \\
64  & 48798  & 67560  & 1852   & 1142   & 5256   & 5092   \\
\bottomrule
\end{tabular}%
}
\label{tab:shoal}
\vspace{-1em}
\end{table}

When using 16 cores, SHOAL's main memory accesses remain high, but shows a significant increase in remote chiplet accesses within the local NUMA. Because it places tasks on 2 chiplets but  frequently moves tasks or data between them, leading to increased overhead of inter-chiplet communication. ARCAS, on the other hand, maintains balanced accesses across the  memory hierarchy, with fewer accesses from the main memory  but more from the local L3 caches. 

The performance gap widens at 32 cores. The remote chiplet accesses in SHOAL significantly increase across 4 chiplets due to its unrestricted core/task replacement and data movement. In contrast, ARCAS maintains balanced task and data placement with more local chiplet accesses. At 64 cores, the memory access patterns of ARCAS and SHOAL become similar. This convergence suggests that at very high core counts, both systems adopt comparable schemes for core/task scheduling.

\begin{figure}[t]
	\centering
	\begin{subfigure}[b]{0.45\textwidth}
		\centering
		\includegraphics[width=\textwidth]{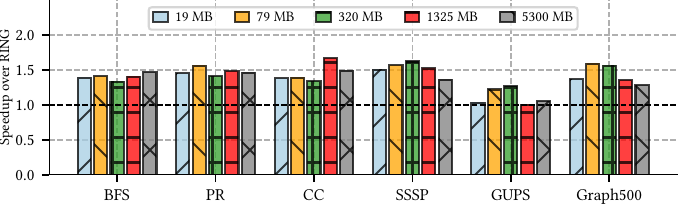}
		\caption{32 Cores}
	\end{subfigure}
	
	\begin{subfigure}[b]{0.45\textwidth}
		\centering
		\includegraphics[width=\textwidth]{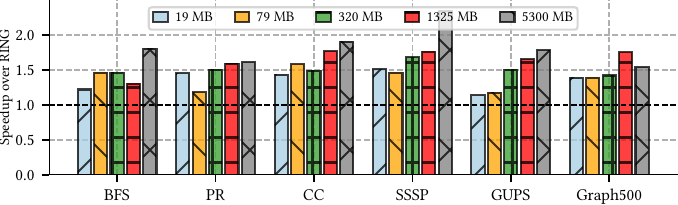}
		\caption{64 Cores}
	\end{subfigure}
	
	\caption{ARCAS's speedups over RING varying the graph size}
	\label{fig:speedupGraphSize}
    \vspace{-0.5cm}
\end{figure}

\subsection{Sensitivity analysis}
\label{sec:eval:SGD}

In this section, we assess the sensitivity of ARCAS's performance with respect
to: (i) the data sizes of workloads and (ii) the irregular access patterns. 

\subsubsection{Data size} 

We first explore the impact of the data size on ARCAS's performance. To this end, we evaluate ARCAS over the 5 graph algorithms and GUPS (see \ref{sec:expSetup}) using different graph configurations, where the dataset size is increased by varying the number of vertices. The datasets range from 19~\unit{MB} with 16 vertices, to 5,300~\unit{MB} with 24 vertices (19~\unit{MB}, 79~\unit{MB}, 320~\unit{MB}, 1,325~\unit{MB}, and 5,300~\unit{MB}). We measure the speedup of ARCAS over the baseline RING using 32 (4 chiplets) and 64 (8 chiplets) cores.


As illustrated in \F\ref{fig:speedupGraphSize}, ARCAS outperforms RING for all 6 benchmarks and core counts. The speedups over RING remain stable with increasing graph sizes, indicating that ARCAS's performance is influenced more by the size of the working set rather than the total data size. Performance remains stable as long as the working set fits within the available cache, regardless of the overall dataset size. With 32 cores, ARCAS shows strong performance at dataset sizes of 79\unit{MB} and 320\unit{MB}, particularly for \emph{SSSP}, \emph{GUPS}, and \emph{Graph500}. These sizes align with the L3 cache capacity of the AMD EPYC Milan processor, allowing ARCAS to effectively optimize the cache utilization.  With 64 cores, ARCAS shows higher speedup over RING with increasing graph sizes, due to RING's limited scalability and ARCAS's balanced placement of cores/tasks across all the chiplets.


ARCAS's adaptive runtime system showcases its superior data placement and task scheduling, the capability to exploit the partitioned L3 caches and the mitigated inter-chiplet data movement. Such flexibility enables ARCAS to maintain high performance across a wide range of graph sizes (within a single chiplet's L3 cache or spanning across multiple chiplets) and core counts. 



\subsubsection{Irregular access patterns}

\begin{figure}[t]
	\centering
	\begin{subfigure}[b]{\columnwidth}
		\centering
		\includegraphics[width=\textwidth]{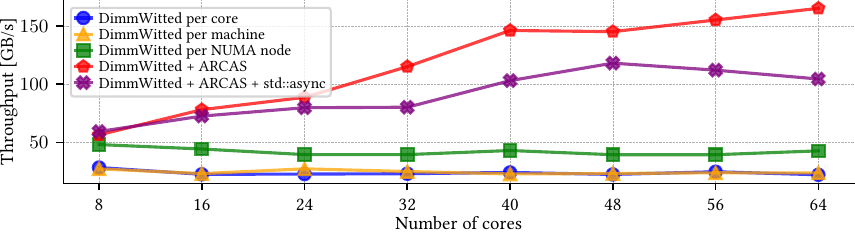}
		\caption{Logistic loss calculation}
		\label{fig:loss_function}
	\end{subfigure}
	
	\begin{subfigure}[b]{\columnwidth}
		\centering
		\includegraphics[width=\textwidth]{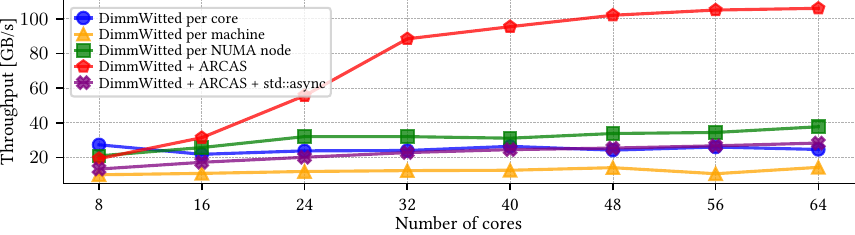}
		\caption{Gradient computation and model update}
		\label{fig:gradient_function}
	\end{subfigure}
	
	\caption{SGD for a logistic regression model}
	\label{fig:dimmwitted}
\end{figure}

Next, we investigate the impact of irregular data / memory access patterns to ARCAS's performance. We evaluate ARCAS using a Stochastic Gradient Descent (SGD) algorithm ~\cite{Carpenter2008LazySS} for logistic regression on a dataset with 10,000 samples and 8,192 features, totaling approximately 6,250\unit{MB}. 
To this end, we have built the ARCAS runtime on top of DimmWitted~\cite{Zhang2014DimmWittedAS}, an analytics engine optimized for statistical computations on modern NUMA architectures (noted as \texttt{\small DimmWitted+ARCAS}). In addition, 
we compare the performance of ARCAS against 4 baselines, including native task scheduling schemes of DimmWitted:
\begin{compactitem}
	\item \texttt{\small DimmWitted-per-core}: assigns each task per CPU core, maximizing the parallelism.
	\item \texttt{\small DimmWitted-NUMA-node}:  assigns each task per NUMA node. It maintains a mutable state for each NUMA node, minimizing inter-node communication.
	\item \texttt{\small DimmWitted-per-machine}: allocates a task for the machine.
\end{compactitem}
\tinyskip
We also use the other baseline
\begin{compactitem}
	\item \texttt{\small DimmWitted+ARCAS+std::async}: employs OS-level task scheduling to replace the coroutines of ARCAS using \texttt{std::async} in standard C++ compiled in GCC 12 with the \texttt{-O3}. 
\end{compactitem}
 We measure the throughput of  the loss function and gradient calculation of the SGD algorithm for \texttt{\small DimmWitted+ARCAS} and all the baselines, ranging the core counts from 8 to 64.

\F\ref{fig:dimmwitted} illustrates the results. 
ARCAS (\ie \texttt{\small DimmWitted\\+ARCAS}) consistently improves performance of DimmWitted with increasing core counts for the logistic loss calculation (\F\ref{fig:loss_function}), with the peaking throughput at 165\unit{GB/s}. \texttt{\small DimmWitted+ARCAS+std::async}  results in a notable drop in throughput, due to the overhead of \texttt{std::async}.
 Among DimmWitted's native strategies, \texttt{\small DimmWitted-NUMA-node} delivers the best performance, reaching 50\unit{GB/s}. However, none of the native strategies scale well with increasing core counts. 
 
The same trend is mirrored at the gradient computation, as demonstrated in \F\ref{fig:gradient_function}. 
The throughput almost remains constant for all DimmWitted native scheduling schemes with increasing core count.   \texttt{\small DimmWitted-NUMA-node} delivers the best performance again, reaching 40\unit{GB/s} at 64 cores. The baseline of \texttt{\small DimmWitted+ARCAS+std::async} performances even worse than \texttt{\small DimmWitted-NUMA-node}, with throughput of 28\unit{GB/s} at 64 cores. ARCAS (\ie \texttt{\small DimmWitted+ARCAS}) significantly boosts the throughput performance reaching up to 106\unit{GB/s} at 64 cores.


\begin{figure}[t]
	\centering
	\begin{subfigure}[b]{0.45\textwidth}
		\centering
		\includegraphics[width=\textwidth]{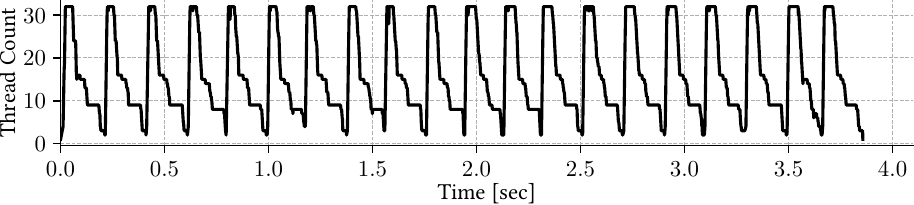}
		\caption{DimmWitted}
        \label{fig:threads_dimmwitted}
	\end{subfigure}
	
	\begin{subfigure}[b]{0.45\textwidth}
		\centering
		\includegraphics[width=\textwidth]{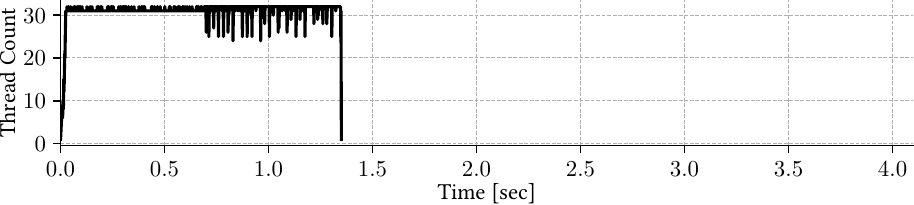}
		\caption{ARCAS}
        \label{fig:thread_charm}
	\end{subfigure}
    \caption{Thread concurrency during SGD with 32 cores and 10,000 exponents.}
    \label{fig:threadconcurrency}
\end{figure}

We attribute ARCAS's superior performance to two reasons: (i) the chiplet-aware task placement and (ii) the lightweight coroutine implementation. 
For reason (i), as already analyzed in \S\ref{sec:eval:Graph} and \S\ref{sec:eval:Shoal}, ARCAS's global scheduler enhances data locality within chiplets, optimizing cache usage and minimizing main memory access. 

However, we would like to highlight that a significant performance boost comes from reason (ii)-- ARCAS's use of coroutines, especially for the gradient computation and model update. Unlike DimmWitted’s reliance on \texttt{std::async} by mapping each task to a separate thread, ARCAS's coroutines run multiple tasks on a single thread and, thereby, significantly reducing the overhead. \F\ref{fig:threadconcurrency} shows the thread concurrency during SGD execution with and without ARCAS.  In particular, DimmWitted presents an average thread count of 16.23 and it fluctuates consistently, resulting in overheads from context switching and poor synchronization.  In contrast, \F\ref{fig:thread_charm} shows a stable thread count of ARCAS, with an average of 31.16 due to its  controlled manner of concurrency management.

 
 

The main limitation of \texttt{std::async} is that it blocks threads, often requiring the creation of more threads to manage tasks. In contrast, ARCAS uses cooperative multitasking, allowing coroutines to yield without blocking, thus running multiple tasks on the same thread. For example, while DimmWitted created 641 threads on 32 cores, ARCAS used only 34, reducing thread creation overhead. Additionally, \texttt{std::async} relies on OS-level thread switching, which is slower than ARCAS’s lightweight user-space context switching, enabling faster task execution.



\subsection{Performance on OLAP}
\label{sec:eval:OLAP}

In this section, we examine the efficacy of ARCAS's runtime over online analytical processing (OLAP) workloads. To this end, we incorporate ARCAS into the DuckDB~\cite{Raasveldt2019DuckDBAE}, overriding the task scheduling and thread mapping management. We evaluate both DuckDB with and without ARCAS (noted as \texttt{\small DuckDB+ARCAS} and  \texttt{\small DuckDB}) using the TPC-H benchmark with a scale factor of 100 using 8 cores (\ie the number of cores in a single chiplet). 

\begin{figure}[t]
    \centering
    \includegraphics[width=\columnwidth]{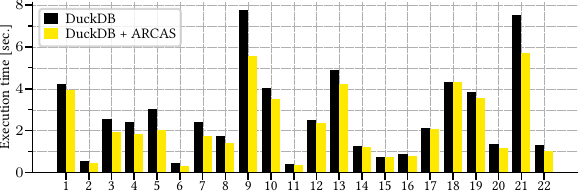}
    \caption{TPC-H queries on DuckDB}
    \label{fig:duckdb}
\end{figure}

\F\ref{fig:duckdb} shows the execution times of each TPC-H query. All the queries show performance improvements (\ie reduced execution time) with ARCAS's chiplet-aware runtime. This reflects that ARCAS introduces negligible runtime overhead. In particular, queries with \texttt{hash-joins} and \texttt{inner-joins} on large tables (\eg Q3, Q4, Q5, Q7, Q9, and Q10), exhibited the most substantial improvements. For example, queries joining the \texttt{lineitem} and \texttt{orders} tables 
benefit significantly with speedups between 1.24$\times$ and 1.51$\times$. 
In particular, Q21 showed a significant spedup of 1.32$\times$ with multiple hash join operations on the \texttt{lineitem} table (e.g., \texttt{\small l\_orderkey=l\_orderkey and l\_suppkey!=l\_suppkey}). Here, the adaptive controller ensures that threads repeatedly accessing the same data remain within the same chiplet, reducing inter-chiplet traffic and improving cache reuse.

The performance gain comes from ARCAS's chiplet-aware adaptive controller that distributes threads across multiple chiplets and fully utilizes the aggregated L3 caches  when executing hash join operations. It reduces cache contention and enhances data locality when scanning large tables (this also results in the speedups of Q8, Q19, and Q20).  On the other hand,  for queries with smaller working sets (\eg Q1, Q2, Q6, and Q11 ), ARCAS's adaptive controller compacts threads within fewer chiplets, reducing inter-chiplet communication and ensuring that data remained closer to the executing threads, which increase cache locality and reducing latency. However, queries with \texttt{hash group-by} operations, \eg Q18, present limited speedups. This is because the data distribution tends to be uneven across chiplets, making it harder to optimize cache utilization.

In conclusion, DuckDB's default thread assignment is chiplet-agnostic, potentially spreading threads across chiplets. ARCAS, however, employs query-specific policies: distributing threads for join-heavy queries and consolidating them for smaller working sets. As for scalability, increasing the thread count would likely narrow the performance gap between \texttt{\small DuckDB} and \texttt{\small DuckDB+ARCAS}, as DuckDB's random distribution would eventually utilize all cores across chiplets.

\subsection{Performance on OLTP}
\label{sec:eval:OLTP}

Finally, we investigate the impact of ARCAS's chiplet-aware scheduling on database performance, \ie online transaction processing (OLTP) workloads. To this end, we have modified ERMIA~\cite{Kim2016ERMIAFM}, a memory-optimized OLTP system, to examine the trade-off of the CHAMR's runtime---between maximizing cache locality and increasing available cache size across multiple chiplets. In particular, we have adapted ERMIA's scheduling schemes to two distinct policies: \texttt{LocalCache} and \texttt{DistributedCache}. 

\texttt{LocalCache} improves cache locality by limiting operations to cores within a few chiplets, reducing inter-chiplet communication but restricting L3 cache size. In contrast, \texttt{DistributedCache} spreads operations across more chiplets, increasing cache capacity but with higher communication overhead. These static policies approximate ARCAS's dynamic task mapping, enabling us to assess the benefits of chiplet-aware scheduling without major changes to ERMIA's architecture. We evaluate their performance at different core counts using the YCSB~\cite{Cooper2010BenchmarkingCS} and TPC-C~\cite{tpcc2010} benchmarks.

We hypothesize that the frequent commits and synchronizations typical of OLTP workloads will dampen the expected performance differences between these two policies. While \texttt{LocalCache} may reduce inter-chiplet communication, the short-lived nature of OLTP transactions and their constant need for synchronization are likely to minimize the benefits of enhanced cache locality. Similarly, \texttt{DistributedCache} may not fully capitalize on its increased cache capacity, as the small data footprint of these transactions reduces the need for larger caches.

\begin{figure}[tb]
     \centering
     \begin{subfigure}[t]{0.9\columnwidth}
        \includegraphics[width=\linewidth]{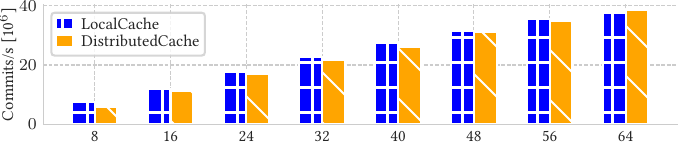}
        \caption{YCSB}
        \label{fig:ycsb}
     \end{subfigure}
     \quad
     \begin{subfigure}[t]{0.9\columnwidth}
        \includegraphics[width=\linewidth]{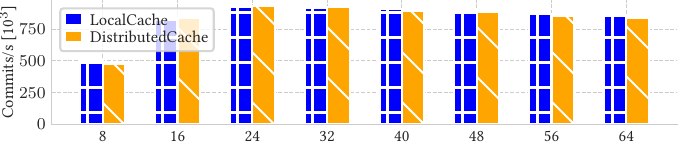}
        \caption{TPC-C}
        \label{fig:tpcc}
     \end{subfigure}
     \caption{Commit per seconds for various scheduling policies.}
     \vspace{-0.5cm}
     \label{fig:OLTP}
\end{figure} 

The YCSB workload, with its simple 45\% read and 55\% read-modify-write operations on a single table, is well-suited for assessing how frequent read-modify-write operations influence cache locality and communication overhead. In contrast, TPC-C's more complex mix of transactional operations and cross-partition accesses provides a scenario to evaluate whether increased cache size can benefit workloads with more diverse access patterns, despite the additional communication overhead.


%

As illustrated in \F\ref{fig:OLTP},  both YCSB (\F\ref{fig:ycsb}) and TPC-C (\F\ref{fig:tpcc}) show nearly identical performance between \texttt{Local\\Cache} and \texttt{DistributedCache} scheduling policies across all core counts. This is because OLTP workloads, characterized by short transactions with frequent commits and synchronizations, are less affected by cache locality or larger aggregated cache sizes. Instead, OLTP performance is often limited by commit latency, synchronization overhead, and maintaining ACID properties, where frequent inter-thread communication and disk I/O overshadow any cache optimization benefits.



%% file: 7-Related_Work/Related-work.tex
\section{RELATED WORK}

\textbf{Cache Partitioning.}
 has been widely studied as a means to reduce interference between applications sharing the LLC. Intel's CAT \cite{intelCAT} provides hardware support for partitioning the LLC. Several works have leveraged it to improve performance isolation and quality of service \cite{Selfa2017ApplicationCP, Xiang2018DCAPSDC}. However, these approaches typically use static partitioning schemes. More dynamic approaches have also been proposed. Qureshi and Patt \cite{Qureshi2006UtilityBasedCP} introduced Utility-Based Cache Partitioning (UCP) that monitors the utility of cache space for each application and partitions the cache accordingly. Jaleel et al. \cite{TADIP} proposed Thread-Aware Dynamic Insertion Policy (TADIP) that dynamically adjusts the cache insertion policy based on thread behavior. While effective, these techniques do not consider the underlying chiplet architecture.


\textbf{Cache-aware scheduling.} Anderson et al. explored cache-aware task scheduling for real-time systems, focusing on partitioning caches to reduce interference between tasks in multicore environments \cite{Guan2009CacheawareSA} while Gracioli et al. reviewed cache management techniques in real-time embedded systems \cite{Gracioli2015ASO}. Recent work has also explored runtime support for emerging chiplet-based architectures. Chirkov et al. evaluated interconnect performance and introduced Meduza, a write-update coherence protocol for chiplet systems \cite{Chirkov2023SeizingTB}. 

\textbf{Adaptive scheduling}. Some prior work already explored the benefits of adaptive task scheduling and data placement, however most of it was in the context of NUMA. For example, OmpSs is a dynamic scheduling approach to efficiently schedule dependent tasks on heterogeneous multi-core systems by prioritizing critical tasks for fast cores \cite{Chronaki2015CriticalityAwareDT}. Psaroudakis et al. proposed an adaptive data placement algorithm that can track utilization imbalances across socket \cite{Psaroudakis2016AdaptiveND}. ATraPos uses dynamic repartitioning for OLTP workloads, to avoid transactions crossing partitions and avoid inter-partition synchronization \cite{Porobic2014ATraPosAT}. ERIS~\cite{Eris} is an in-memory database engine that uses adaptive partitioning to mitigate synchronization overheads and improve performance.

\textbf{Chiplet-awareness.} 
The notable heterogeneity of inter-core latencies and the memory hierarchy has also been documented by Velten et al.~\cite{Velten2022MemoryPO}. Additional studies on chiplet performance include Suggs et al.'s examination of AMD’s Zen 2 architecture \cite{ZEN2}, Schöne et al.'s work on improving energy efficiency \cite{ZEN2energy}, and Naffziger et al.'s analysis of multi-die configurations \cite{2.2AMD}.
Fogli et al. explored the impact of chiplet-based architectures on the performance of OLAP workloads and proposed deployment strategies to optimize distributed query execution~\cite{Fogli2024OLAPOM}. Another work proposes concrete optimization steps for a high-performance sorting algorithm in the context of chiplets~\cite{FogliSorting2024}.

%% file: 8-Conclusion/Conclusion.tex
\section{CONCLUSION}

Our research highlights the impact that chiplet-based CPU architectures can have on parallel processing and data-intensive applications. We introduced ARCAS, a runtime system that optimizes task allocation and resource management across chiplets. Through extensive experiments, we show that ARCAS can outperform state-of-the-art NUMA-aware systems, achieving up to $3.85\times$ speedup in graph processing and maintaining superior performance for other memory-intensive workloads.

Our findings are especially beneficial for workloads with irregular access patterns, customizing their resource allocation policies on novel chiplet-based processors. This research work underscores the need for data-intensive systems to move beyond conventional NUMA-aware optimizations, embracing new task scheduling approaches that fully exploit the capabilities of modern chiplet-based processors.